\documentclass[11pt]{article}
\usepackage{natbib, graphicx, setspace, amsmath, amssymb, subfigure, url, multirow, booktabs, verbatim, bm,threeparttable,algorithm, color}
\usepackage{amsthm}
\usepackage{natbib}

\allowdisplaybreaks[3]
\begin{document}
	
	\setlength{\textheight}{575pt}
	\setlength{\baselineskip}{23pt}
	
	\title{Gene-gene interaction analysis incorporating network information via a structured Bayesian approach} 
	\author{Xing Qin$^{1}$,
		Shuangge Ma$^{2}$, and Mengyun Wu$^{1,*}$\\ \\
		$^{1}$	School of Statistics and Management, Shanghai University of Finance and Economics \\
		$^{2}$Department of Biostatistics, Yale School of Public Health\\
	\\
	email: wu.mengyun@mail.shufe.edu.cn}
	\date{} 
	\maketitle

\begin{abstract}{
		Increasing evidence has shown that gene-gene interactions have important effects on biological processes of human diseases. Due to the high dimensionality of genetic measurements, existing interaction analysis methods usually suffer from a lack of sufficient information and are still unsatisfactory. Biological networks have been massively accumulated, allowing researchers to identify biomarkers from a system perspective by utilizing network selection (consisting of functionally related biomarkers) as well as network structures. In the main-effect analysis, network information has been widely incorporated. However, there is still a big gap in the context of interaction analysis. Recently, link networks describing the relationship between genetic interactions have been demonstrated to effectively reveal multi-scale hierarchical organisation in networks and provide interesting findings beyond node networks. In this study, we develop a novel structured Bayesian interaction analysis approach, effectively incorporating the network information. This study is among the first to identify gene-gene interactions with the assistance of network selection for phenotype prediction, while simultaneously accommodating the underlying network structures of both main effects and interactions. It innovatively respects the multiple hierarchies among main effects, interactions, and networks. Bayesian method is adopted, which has been shown to have multiple advantages over some other techniques. An efficient variational Bayesian expectation-maximization algorithm is developed to explore the posterior distribution. Extensive simulation studies demonstrate the practical superiority of the proposed approach. The analysis of TCGA data on melanoma and lung cancer leads to biologically sensible findings with satisfactory prediction accuracy and selection stability. }
	{Assistance of network selection; Gene-gene interaction; Link network; Structured analysis.}
\end{abstract}

\maketitle

\section{Introduction}
\label{s:intro}

Gene-gene interactions have  significant importance for the basis of human diseases beyond main genetic effects \citep{Intro2009detecting,Intro2014epistasis}. Due to the higher dimensionality, lower signal-to-noise ratio, and other reasons, there are more challenges in the analysis of interactions  compared to main effects. We refer to \cite{Intro2016review}, \cite{Intro2019robust}, and the references therein for more discussions. In recent interaction analysis research, the ``main effects-interactions'' hierarchy is generally employed to improve both estimation and interpretation \citep{Hao2017A}. Specifically, an interaction can be identified only when one of its main effects (weak hierarchy) or both (strong hierarchy) are also identified. A number of statistical methods have been developed to identify important interactions and enforce this hierarchy. Among the available techniques, penalization has drawn much attention. Published works include the Lasso for hierarchical interaction \citep{Intro2013a1}, interaction learning via a hierarchical group-lasso regularization \citep{Glin2015learning}, penalized tensor regression \citep{Introwu2018identifying}, and quadratic regression under the marginality principle \citep{hao2018model}.

Despite the vast literature on penalization and some other methods, there are very few Bayesian methods for hierarchical interaction analysis. Limited existing studies include \cite{Intro2015bayesian1}, which proposes a Bayesian hierarchical mixture model for interaction analysis and incorporates the natural hierarchical structure using the conditional prior probability technique. As another example, \cite{kim2018bayesian} develops a Bayesian interaction analysis method with a hierarchical prior that fully considers the hierarchy constraint and controls the degree of sparsity simultaneously. There are also a few recent Bayesian methodological developments without enforcing hierarchy, including \cite{ren2020semiparametric} and \cite{ferrari2020bayesian}.

With the high dimensions of genetic measurements but still limited sample sizes, the existing interaction analysis usually suffers from a lack of sufficient information and leads to unsatisfactory results. To improve identification and predictive performance, in main-effect analysis, a promising direction is to incorporate biological network information, which can be roughly classified into two strategies. The first strategy has been developed to take advantage of the assistance of network selection, where the ``main effects-networks'' hierarchy is usually enforced. That is, a main effect can be included in the model only when at least one of its involved networks is also included, and vice versa. Examples include a bi-level selection approach using the group exponential Lasso \citep{Intro2015GEL}, and the Bayesian sparse group selection with spike and slab priors  \citep{xu2015bayesian}. Complementary to the first strategy, the second strategy has been developed to incorporate network structures. A representative technique is the network regularization based on the graph Laplacian matrix. Examples include penalization methods with the Laplacian-based penalty \citep{Intro2008LL1,Intro2019Liu}, and
Bayesian methods with the Laplacian Gaussian prior \citep{Intro2020laplace}. To take advantages of both strategies, multiple Bayesian methods have been developed to utilize network selection and also effectively account for underlying network structures \citep{Intro2013joint,Intro2014a,Intro2016joint}. However, most existing methods have been designed for main-effect analysis, and methodological developments in the context of interaction analysis are still very limited.

As stated in \cite{0Link}, beyond the traditional networks with nodes being the genetic factors, link networks describing the relationship between genetic interactions can effectively reveal multi-scale hierarchical organisation in networks. The identified link networks based on, for example the protein-protein interaction and metabolic networks, have been shown to have important biological implication in \cite{0Link}. They may contribute to predict more detailed and interpretable roles of oncogenes \citep{ahn2011integrative}, reveal cell functional organization and potential cellular mechanisms \citep{wang2014functional}, determine whether the drug action area is part of the protein-interaction interface \citep{dos2018building}, and others. Recent successes of incorporating network information in the main-effect analysis and the importance of link networks call for the effective network integration approaches for interaction analysis.

In this study, we propose a new structured Bayesian interaction analysis approach. This study is the first to conduct gene-gene interaction analysis with the assistance of network selection and simultaneously accommodate network structures. The most significant advancement is that both the ``main effects-interactions'' and ``main effects/interactions-networks'' hierarchy conditions are effectively respected, which is much more challenging than in the existing interaction analysis or network selection-assisted main effect analysis that enforce only one hierarchy. Furthermore, the underlying network structures are explored in the analysis of not only main effects but also interactions, making this study a big step forward from the existing main effect structured analysis. The proposed approach is based on the Bayesian method, which has been shown to have multiple advantages over some other techniques, such as penalization \citep{gibbs}. Different from most published Bayesian interaction studies based on the Markov Chain Monte Carlo (MCMC) inference technique, we take advantage of the hybrid model integrating conditional and generative components, and develop a more efficient variational Bayesian expectation-maximization algorithm. This is especially desirable with the extremely high dimensions in gene-gene interaction analysis. Overall, this study can provide a useful new venue for genetic interaction analysis.

\section{Methods}

Consider $K$ networks $G_1\left({V}_1,{E}_1\right),\cdots,G_K\left(V_K, E_K\right)$, which have been constructed using the existing biological network information. Here $V_k$ is the node set consisting of $p_k$ genetic factors, and ${E}_k=\left(e_k\left(j,l\right)\right)_{p_k\times p_k}$ is the set of edges between two nodes. Suppose that we have $n$ i.i.d. subjects with ${\boldsymbol{X}}= \left(\boldsymbol{X}_1,\cdots,\boldsymbol{X}_p\right) \in \mathrm{R}^{n \times p}$  being the matrix of all genetic measurements, and  $\boldsymbol{y}\in \mathrm{R}^{n \times 1}$ being the response vector, where  $\boldsymbol{X}_j$ is a $n\times 1$ vector for $j=1,\cdots,p$, and $p=\sum_{k=1}^K p_k$.
Note that if a genetic factor is involved in multiple networks, the corresponding measurement is duplicated in these networks.

\subsection{Model}
We consider the most popular continuous response, and the proposed approach can be extended to other responses. Specifically, consider the following linear model:
\begin{equation}\label{equation:model}
	\boldsymbol{y}
	=\sum_{j=1}^p w_j^{(\boldsymbol{1})}\boldsymbol{X}_j+\sum_{l_1=1}^p \sum_{l_2>l_1}^p w_{l_1l_2}^{(\boldsymbol{2})}\boldsymbol{X}_{l_1}\circ \boldsymbol{X}_{l_2}+\boldsymbol{\epsilon}\triangleq\tilde{\boldsymbol{X}}\boldsymbol{w}+\boldsymbol{\epsilon},
\end{equation}
where  $\circ$ denotes the element-wise product,  $\tilde{\boldsymbol{X}}\in \mathrm{R}^{n \times \left(p(p+1)/2\right)}$ is the matrix of all genetic measurements $\boldsymbol{X}_j$ and their interactions $\boldsymbol{X}_{l_1}\circ \boldsymbol{X}_{l_2}$,
$\boldsymbol{w}=\left(w_1^{(\boldsymbol{1})},\cdots,w_p^{(\boldsymbol{1})},w_{12}^{(\boldsymbol{2})},\cdots,w_{(p-1)p}^{(\boldsymbol{2})}\right)^{\mathrm{T}}\triangleq\left(w_{j}\right)_{\left(p(p+1)/2\right) \times 1}$, and
$\boldsymbol{\epsilon}\sim\mathcal{N}\left( 0, \tau^{-1}\boldsymbol{I}\right)$ with $ \boldsymbol{I}$ being an identity matrix and $\tau$ being a precision parameter.

First, to accommodate network structure for main effects $x_{j}$'s ($x_j$: the $j$th factor corresponding to $\boldsymbol{X}_j$),  for the $k$th network $G_k(V_k, E_k)$, an adjacency matrix $\boldsymbol{A}_k^{(\boldsymbol{1})}$ is constructed, where $A^{(\boldsymbol{1})}_{k}(j,l)=1$ if there is an edge $e_k\left(j,l\right)$ between the $j$th and $l$th factors, and $A^{(\boldsymbol{1})}_{k}(j,l)=0$ otherwise. In addition, for the interactions $x_{j_1} x_{l_1}$ and $x_{j_2} x_{l_2}$ ($l_1\neq l_2$) of which the corresponding main effects are involved in the $k$th network $G_k$,  we construct a line graph following \cite{0Link}.  Specifically,  a similarity is first defined as
$$S_{k}\left(x_{j_1}x_{l_1},x_{j_2}x_{l_2}\right)=\begin{cases}{\frac{\left|n_{+}(l_1) \cap n_{+}(l_2)\right|}{\left|n_{+}(l_1) \cup n_{+}(l_2)\right|},}&{\text{~if~} j_1=j_2, \text{~that is, the interactions share a main effect~} x_{j_1},
	}\\
	{	0,}&{\text{~otherwise}},
\end{cases}$$
where  $n_{+}(l)$ denotes the set of the main effect $x_l$ and its neighbors with edges in $G_k$. The adjacency matrix  $\boldsymbol{A}^{(\boldsymbol{2})}_k$ for the interactions is constructed with  $A^{(\boldsymbol{2})}_{k}\left(x_{j_1}x_{l_1},x_{j_2}x_{l_2}\right)=\boldsymbol{1}_{\left\{S_{k}\left(x_{j_1}x_{l_1},x_{j_2}x_{l_2}\right)>0\right\}}$, where $\boldsymbol{1}_{\{\cdot\}}$ is an indicator function. A toy example on the network construction of interactions is provided in Figure 1.
\begin{figure}[h]
	\centering
	\includegraphics[width=5.8in]{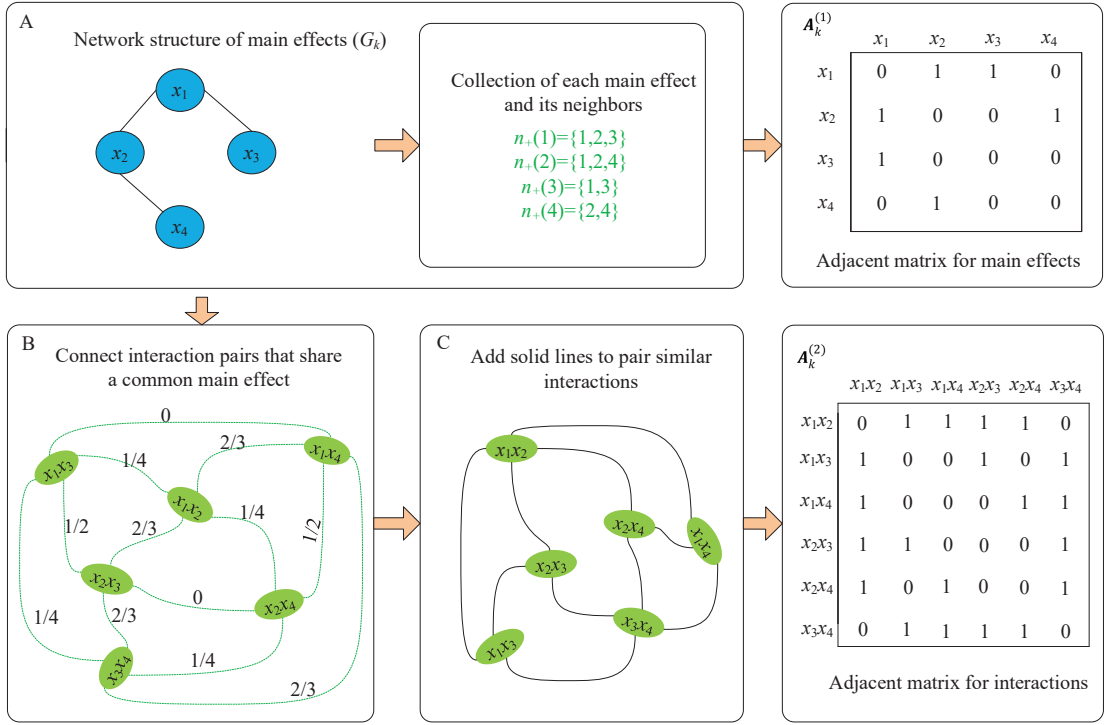}
	\caption{A toy example on the network construction of interactions. A: The network for main effects. B: Establish relationships and calculate pairwise similarity defined as $S_{k}\left(x_{j_1}x_{l_1},x_{j_2}x_{l_2}\right)= \frac{\left|n_{+}(l_1) \cap n_{+}(l_2)\right|}{\left|n_{+}(l_1) \cup n_{+}(l_2)\right|}\boldsymbol{1}_{\left\{j_1=j_2\right\}}$ for interactions. Here the interaction pairs that share a common main effect are connected with a dashed line, where the corresponding value of $S_k(\cdot,\cdot)$ is provided.  For example, the similarity between interactions $x_{1}x_{3}$ and $x_{1}x_{4}$ is 0 ($n_{+}(3)=\left\{1,3\right\},n_{+}(4)=\left\{2,4\right\},\text{~so~}S_{k}\left(x_{1}x_{3},x_{1}x_{4}\right)= \frac{\left|n_{+}(3) \cap n_{+}(4)\right|}{\left|n_{+}(3) \cup n_{+}(4)\right|}=0$). C: Construction of the network for the interactions, where there is an edge (solid line) between two interactions if $ S_{k}\left(x_{j_1}x_{l_1},x_{j_2}x_{l_2}\right)>0 $. }
	\label{fig:networks}
\end{figure}
Then, the hierarchical  representation for the proposed model is:
\begin{equation}\label{equation:prior}
	\begin{aligned}
		&\boldsymbol{y}|\boldsymbol{w}\sim\mathcal{N}\left(\tilde{\boldsymbol{X}}\boldsymbol{w},\tau^{-1}\boldsymbol{I}\right),
		{{w}}_{j} |{\beta}_{j}\sim \mathcal{N}\left(0, s_{1}\right)^{\beta_{j} } \mathcal{N}\left(0, s_{2}\right)^{1-\beta_{j} }, \beta_{j}|\zeta_{j} \sim\mathrm{Bern}\left(\zeta_{j}\right), \zeta_{j} \sim\mathrm{Beta}\left(a,b\right), %\triangleq\mathcal{SSN}\left(w_j|0,\beta_j\right),
		\\&	 \bar{{w}}_{l_1 l_2}^{(\boldsymbol{2})}|  {\beta}_{l_1}^{(\boldsymbol{1})}{\beta}_{l_2}^{(\boldsymbol{1})} \sim \mathcal{N}\left(0, s_{1}\right)^{{\beta}_{l_1}^{(\boldsymbol{1})} {\beta}_{l_2}^{(\boldsymbol{1})}} \mathcal{N}\left(0, s_{2}\right)^{1-{\beta}_{l_1}^{(\boldsymbol{1})} {\beta}_{l_2}^{(\boldsymbol{1})}}, \bar{w}_{l_1 l_2}^{(\boldsymbol{2})}|  {w}_{l_1 l_2}^{(\boldsymbol{2})}\sim\boldsymbol{1}_{\left\{\bar  {w}_{l_1 l_2}^{(\boldsymbol{2})}={w}_{l_1 l_2}^{(\boldsymbol{2})}\right\}}, \\&
		\tilde{\boldsymbol{w}}_{k}|\alpha_{k}\sim\mathcal{N}\left(\boldsymbol{0}, s_{1}
		\left({\boldsymbol{L}}_{k}+\xi \boldsymbol{I}\right)^{-1}\right)^{\alpha_{k}} \mathcal{N}\left(\boldsymbol{0}, s_{2} \boldsymbol{I}\right)^{1-\alpha_{k}}, \alpha_{k}\sim\mathrm{Bern}\left(\theta\right), \tilde{\boldsymbol{w}}_k |\boldsymbol{w}_k\sim\boldsymbol{1}_{\{\tilde{\boldsymbol{w}}_k=\boldsymbol{w}_k\}}.
	\end{aligned}
\end{equation}
Here, $\beta_j$ is the selection indicator of $j$th main effect/interaction, with $\beta_{j}=1$ if the $j$th variable is selected and 0 otherwise, and we use $\beta_l^{(\boldsymbol{1})}$ to denote the main-effect-selection indicator corresponding to $w_{l}^{(\boldsymbol{1})}$ for simplicity. $s_{1}$ and $s_{2}$ are two parameters with $s_1>s_2>0$ and $s_{2}$ being very small. $\bar{w}_{l_1l_2}^{(\boldsymbol{2})}$ is a latent variable for $w_{l_1l_2}^{(\boldsymbol{2})}$.  $\tilde{\boldsymbol{w}}_k$ is  a latent vector for ${\boldsymbol{w}}_k=\left\{w_j^{(\boldsymbol{1})} |j \in V_k \right\}\cup\left\{ w_{l_1 l_2}^{(\boldsymbol{2})}| l_1,l_2\in V_k,l_1<l_2\right\}$, which denotes the vector of all regression coefficients in the $k$th network. $\alpha_{k}$ is the network-selection indicator.  ${\boldsymbol{L}}_{k} =\boldsymbol{I}-\boldsymbol{D}_k^{-1/2}\boldsymbol{A}_{k}\boldsymbol{D}_k^{-1/2}$ with $\boldsymbol{A}_{k}=\left(\begin{array}{cc}{\boldsymbol{A}^{(\boldsymbol{1})}_{k}} & {\boldsymbol{0}} \\ {\boldsymbol{0}} & {\boldsymbol{A}_{k}^{(\boldsymbol{2})}}\end{array}\right)$ and  $\boldsymbol{D}_k=\textrm{diag}\left(\sum_{l=1}^{\hat{p}_k}\boldsymbol{A}_{k}(1,l),\cdots, \sum_{l=1}^{\hat{p}_k}\boldsymbol{A}_{k}(\hat{p}_k,l)\right)$, where $\hat{p}_k=p_k(p_k+1)/2$. $\xi$ is a small constant ($\xi=10^{-6}$ in our numerical studies) to make ${\boldsymbol{L}}_{k}+\xi \boldsymbol{I}$ strictly positive-definite. The graphical representation of (\ref{equation:prior}) and detailed posterior computations are given in Section S1 of the Supplementary material. Denote $\textrm{E}(\beta_{j})$ and $\textrm{E}(\alpha_{k})$ as the corresponding posterior expectations of the selection indicators. Following \cite{gibbs}, we adopt the thresholding approach, the main effects (interactions) with $\textrm{E}(\beta_{j})$'s and networks with $\textrm{E}(\alpha_{k})$'s larger than 0.5 are identified as important.

The proposed model has been motivated by the following considerations. Identification of main effects and interactions is achieved using the spike and slab prior $\mathcal{N}\left(0, s_{1}\right)^{\beta_{j} } \mathcal{N}\left(0, s_{2}\right)^{1-\beta_{j} }$, where $\beta_j=0$ leads to the spike component related to $s_2$ with a very small value, so $w_j$ will be truncated to zero. $\bar{{w}}_{l_1l_2}^{(\boldsymbol{2})}$  with  prior based on the main-effect-selection indicators $\beta_{l_1}^{(\boldsymbol{1})}$ and $\beta_{l_2}^{(\boldsymbol{1})}$, together with the indicator function $\boldsymbol{1}_{\left\{\bar  {w}_{l_1 l_2}^{(\boldsymbol{2})}={w}_{l_1 l_2}^{(\boldsymbol{2})}\right\}}$ is developed to accommodate the strong ``main effects-interactions'' hierarchy. Specifically,  if an interaction is selected with $\beta_{l_1l_2}^{(\boldsymbol{2})}=1$, then $\bar{w}_{l_1l_2}^{(\boldsymbol{2})}={w}_{l_1l_2}^{(\boldsymbol{2})}\neq 0$, leading to $\beta_{l_1}^{(\boldsymbol{1})}\beta_{l_2}^{(\boldsymbol{1})}=1$ (i.e. $\beta_{l_1}^{(\boldsymbol{1})}=\beta_{l_2}^{(\boldsymbol{1})}=1$) with a higher probability.   $\tilde{\boldsymbol{w}}_{k}$ and $\alpha_k$ are introduced to assist the selection of interactions (main effects) by network identification and also accommodate the network structures of both main effects and interactions. Specifically, when the $k$th network is selected ($\alpha_{k}=1$), the precision matrix for $\tilde{\boldsymbol{w}}_{k}$ is related to the Laplacian matrix ${\boldsymbol{L}}_{k}$, where the $j$th and $l$th variables are conditionally dependent if ${A}_k(j,l)=1$. Therefore, the effects of connected factors over the $k$th network are promoted to be similar. The ``main effects/interactions-networks'' hierarchy is achieved using $\boldsymbol{1}_{\{\tilde{\boldsymbol{w}}_k=\boldsymbol{w}_k\}}$. If $\alpha_{k}=0$, we have $\boldsymbol{w}_{k}=\tilde{\boldsymbol{w}}_{k}\approx\mathbf{0}$, leading to all $\beta_j$'s in the $k$th network being zero with a higher probability. Moreover, if at least one of $\beta_j$'s belonging to the $k$th network is nonzero, $\alpha_{k}$ is also nonzero with a higher probability.

\subsection{Computation}

For computation, we rewrite priors for $\bar{w}_{l_1l_2}^{(\boldsymbol{2})}$ and $\tilde{\boldsymbol{w}}_k$ as the generative models with observation vector $\boldsymbol{0}$. As such, the proposed approach can be formulated as a hybrid Bayesian model which includes tractable partition functions and can be effectively approximated with the variational Bayesian expectation-maximization (EM) algorithm. Compared to MCMC techniques, variational approximation is computationally more efficient and more feasible with high dimensional parameters. Specifically, we consider minimizing the Kullback-Leibler (KL) divergence between the exact and approximate posterior distributions:
$$
\mathrm{KL}\left(q(\Omega)\|p\left(\Omega | \boldsymbol{y}, \boldsymbol{X};\tau,\theta\right)\right)=\int q(\Omega)\log \left[\frac{ q(\Omega)}{p\left(\Omega | \boldsymbol{y}, \boldsymbol{X};\tau,\theta\right)}\right]\, d\Omega,
$$
where $q(\Omega)=q(\boldsymbol{w})q(\boldsymbol{\beta})q(\boldsymbol{\alpha})q(\boldsymbol{\zeta}) $ is a candidate approximate distribution of our true posterior distribution $p\left(\Omega | \boldsymbol{y}, \boldsymbol{X};\tau,\theta\right)$, and $\Omega$ represents all latent variables. Note that with the distributions of $\bar{w}_{l_1 l_2}^{(\boldsymbol{2})}|  {w}_{l_1 l_2}^{(\boldsymbol{2})}$ and $\widetilde{\boldsymbol{w}}_{k} | \boldsymbol{w}_{k}$ being the indicator functions, there is no need to include the separate distributions $q(\widetilde{\boldsymbol{w}})$ and $q(\bar{\boldsymbol{w}})$ in $q(\Omega)$. In the E step, we optimize the  KL divergence with respect to $q(\Omega)$ while holding the model parameters $\left\{\tau,\theta\right\}$ fixed. After some derivations, we obtain the optimal variational distribution $q(\Omega)$ as follows,
$$	q(\boldsymbol{w})=\prod_{j=1}^{p(p+1)/2}\mathcal{N}\left(m_{j} ,  \sigma_{j} ^{2}  \right),q(\boldsymbol{\beta})=\prod_{j=1}^{p(p+1)/2} \eta_{j}^{\beta_{j}}\left(1-\eta_{j}\right)^{1-\beta_{j}},$$ $$q(\boldsymbol{\zeta}) \propto \prod_{j=1}^{ p(p+1)/2}\left(\zeta_{j}\right)^{\tilde{a}_{j}-1}\left(1-\zeta_{j}\right)^{\tilde{b}_{j}-1},q(\boldsymbol{\alpha})=\prod_{k=1}^{K} r_{k}^{\alpha_{k}}\left(1-r_{k}\right)^{1-\alpha_{k}},$$
where  $\left(m_{j}, \sigma_{j}^2\right)$ and $\left(\tilde{a}_{j}, \tilde{b}_{j} \right)$   are the corresponding estimated values of the parameters for the Gaussian and Beta distributions, respectively, and $\eta_{j}$ and $r_{k}$ are the expectation of $\beta_{j}$ and $\alpha_k$ under $q(\Omega)$.  In the M step, we optimize the KL divergence with respect to the model parameters while keeping the variational parameters $\Omega$ fixed. The proposed algorithm iteratively updates the estimators between the E and M steps until convergence and adopts the final estimated values of $\eta_{j}$ and $r_{k}$ as the estimators of $\textrm{E}(\beta_{j})$ and $\textrm{E}(\alpha_{k})$. We refer to the Section S1 and Algorithm 1 of the Supplementary material for the detailed computation.

To proceed with this algorithm, we consider a uniform Beta prior with $a =b=1$ following the literature \citep{Intro2013joint}. The proposed model involves two tuning parameters $s_{1}$ and $s_{2}$. Our numerical investigation suggests that the value of $s_{1}$ is not very important when it is in a sensible range. To reduce computational cost, we fix $s_1=1$ in our numerical studies. The value of $s_{2}$ is selected using the Bayesian information criterion (BIC). The proposed algorithm is computationally feasible. Take a simulated dataset with $p=1,000$ and $n=300$ as an example. With fixed tuning parameter, the proposed analysis takes about half a minute using a laptop with standard configurations. To facilitate data analysis, we have developed R package \textit{JNNI} implementing the proposed approach, which is publicly available at https://github.com/mengyunwu2020/JNNI and can be installed with devtools.

\section{Simulation}
We perform simulations to evaluate performance of the proposed approach under the following settings. (a) $n=300$ and $p=1,000$. Thus, there are a total of $1,000$ candidate main effects and 499,500 interactions. (b) Consider two settings for the number of networks with $K=100$ and $50$. (c) We follow the network construction procedure used in \cite{zhao2016a}. Specifically, for the $k$th network ($k=1, \cdots, K$), set $p_k=\frac{p}{K}$, generate one transcription factor (TF) $x_{TF}$ from $\mathcal{N} \left(0, 1 \right)$, and then generate the rest $p_k-1$ genetic factors from $\mathcal{N}\left(\rho x_{TF}, 1-\rho^2\right)$ with a parameter $\rho$. Consider $\rho= 0.4$ and $0.6$, representing  different dependence between the TF and its target factors in each network. Genetic factors with nonzero correlations are connected in the network. (d) There are three important networks, where 18 main genetic effects and 17 interactions have nonzero coefficients. Both the ``main effects/interactions-networks'' and ``main effects-interactions'' hierarchies are satisfied. Nonzero signals of the important TFs are generated from Uniform(0.8,1.2), and the other important main effects and interactions have relatively weaker signals with a ratio $r$ of that of the corresponding important TF. Consider $r=1/\sqrt{5}$ and $1/\sqrt{12}$. Four specific settings S1-S4 for the important variables are considered. Under setting S1, all signals are positive. Setting S2 is the same as S1, except that the signals in the second network and those between the first and second networks are negative. Under setting S3, within each network, the signals can be either positive or negative. Under setting S4, the important interactions only involve the none-TF main effects with weaker signals. We refer to the Section S2 of the Supplementary material for more details. (e) For the response, we generate $\boldsymbol{y}$ from the Gaussian distribution (\ref{equation:model}) with variance 1. There are 32 scenarios, comprehensively covering a wide spectrum with different levels of correlations within networks and signals associated with the response, and different patterns of networks and associations.

In addition to the proposed approach, six alternatives are conducted. (a) glinternet, which learns linear interaction model based on the hierarchical group-Lasso regularization and is implemented using the R package \textit{glinternet} \citep{Glin2015learning}. (b) Lasso, which applies the Lasso penalization to both main effects and all pairwise interactions directly and is realized using the R package \textit{glmnet}. (c) iFORM, which identifies interactions in a greedy forward fashion while maintaining the hierarchical structure \citep{Ifort}. (d) HierNet, which is  Lasso for hierarchical interactions by adding a set of convex constraints and is realized using the R package \textit{HierNet} \citep{Intro2013a1}. (e) Grace, which applies the graph-constrained estimation method developed by \cite{Intro2008LL1} to both main effects and all pairwise interactions. (f) GEL, which achieves a bi-level variable selection for groups and individual predictors (main effects and interactions) in those groups \citep{Intro2015GEL}. Among these methods,  glinternet and iFORM respect the strong ``main effects-interactions'' hierarchy. We consider HierNet with the weak hierarchy, as the counterpart with strong hierarchy is not computationally feasible in large-scale simulations. Lasso, Grace, and GEL are originally developed for main-effect analysis, and we extend them for interaction analysis by modeling additional all pairwise interactions, without enforcing the ``main effects-interactions'' hierarchy. Both Grace and GEL incorporate the network information, where Grace accommodates the underlying network structures, and GEL  achieves the joint selection of interactions and networks.

To evaluate identification performance, we compute the numbers of  true positives and false positives for main effects (M:TP and M:FP) and interactions (I:TP and I:FP), respectively. For the proposed approach and GEL, we also consider the true positives and false positives (N:TP and N:FP) for identifying networks. Estimation performance is assessed using the root sum of squared errors (RSSE) defined as $||\hat{\boldsymbol{w}}_{\mathcal{M}}-\boldsymbol{w}_{\mathcal{M}}^{0}||_{2}$ and $||\hat{\boldsymbol{w}}_{\mathcal{I}}-\boldsymbol{w}_{\mathcal{I}}^{0}||_{2}$ for  main effects and interactions, where $(\hat{\boldsymbol{w}}_{\mathcal{M}},\hat{\boldsymbol{w}}_{\mathcal{I}})$ and $(\boldsymbol{w}_{\mathcal{M}}^{0},\boldsymbol{w}_{\mathcal{I}}^{0})$ are the estimated and true values of coefficients. For prediction evaluation, we adopt the prediction median-squared error (PMSE) based on independent testing data with 100 subjects.

Under each scenario, we simulate 100 replications. The summary results under the scenarios with $\rho=0.4$ and $K=100$ are presented in Table 1 ($r=1/\sqrt{5})$ and Table 2 ($r=1/\sqrt{12}$). The rest of the results are shown in the Section S2 of the Supplementary material.

\begin{table}\label{t:one}
	\renewcommand\tabcolsep{1.4pt}
	\caption{Simulation results under the scenarios with $\rho=0.4, K=100$, and $r=1/\sqrt{5}$. In each cell, mean (SD) based on 100 replicates.}
	\begin{tabular}{@{}lccccccc@{}}
		\toprule
		\textbf{Approach} & \textbf{M:TP} & \textbf{M:FP} & \textbf{M:RSSE} & \textbf{I:TP} & \textbf{I:FP} & \textbf{I:RSSE} & \textbf{PMSE} \\ \midrule
		\multicolumn{8}{c}{S1}                                                                                                           \\
		proposed       & 17.86(0.35)& 	2.06(1.06)& 	0.35(0.10)& 	16.90(0.30)	& 9.44(3.64)& 	0.45(0.09)& 	0.55(0.10)
		\\
		glinternet      & 17.34(1.02)   & 5.60(3.02)    & 0.96(0.13)      & 14.42(1.96)   & 4.84(3.01)    & 1.02(0.13)      & 1.51(0.54)    \\
		Lasso         & 8.16(2.62)    & 0.00(0.00)    & 1.48(0.10)      & 11.50(1.96)   & 14.82(6.52)   & 1.19(0.12)      & 2.32(0.71)    \\
		iFORM           & 16.58(1.96)   & 37.38(4.26)   & 1.35(0.31)      & 12.58(3.84)   & 30.04(3.81)   & 1.32(0.34)      & 2.25(1.45)    \\
		HierNet         & 13.72(2.65)   & 1.00(1.43)    & 1.23(0.16)      & 8.76(2.70)    & 8.28(6.06)    & 1.32(0.12)      & 2.31(0.71)    \\
		Grace              & 9.50(2.53)    & 0.38(0.88)    & 1.46(0.13)      & 12.00(1.73)   & 7.02(7.08)    & 1.20(0.09)      & 2.26(0.59)    \\
		GEL             & 17.76(1.17)   & 9.14(4.90)    & 0.76(0.16)      & 12.70(1.47)   & 104.60(53.00) & 1.93(0.51)      & 1.81(0.74)    \\ \midrule
		\multicolumn{8}{c}{S2}                                                                             \\
		proposed       &   17.30(0.79)&	1.77(1.41)&	0.43(0.13)&	16.57(0.68)	&8.17(3.56)&	0.52(0.12)&	0.69(0.18)  \\
		glinternet      & 17.20(1.03)   & 5.68(3.62)    & 0.98(0.12)      & 14.36(1.70)   & 4.50(2.31)    & 1.06(0.13)      & 1.48(0.44)    \\
		Lasso         & 6.48(2.76)    & 0.20(0.64)    & 1.49(0.09)      & 7.86(3.51)    & 50.80(103.98) & 1.44(0.10)      & 2.91(0.71)    \\
		iFORM           & 15.52(2.76)   & 38.92(4.81)   & 1.49(0.43)      & 11.76(4.52)   & 30.56(4.12)   & 1.44(0.42)      & 2.28(1.17)    \\
		HierNet         & 12.42(3.94)   & 1.26(1.58)    & 1.28(0.19)      & 8.62(2.80)    & 8.04(5.97)    & 1.39(0.13)      & 2.15(0.71)    \\
		Grace              & 5.52(1.61)    & 0.20(0.49)    & 1.73(0.05)      & 4.88(1.38)    & 7.62(6.81)    & 1.50(0.05)      & 3.49(0.95)    \\
		GEL             & 17.78(1.09)   & 11.30(2.53)   & 0.78(0.16)      & 13.08(0.72)   & 127.00(34.46) & 2.06(0.34)      & 1.86(0.56)    \\ \midrule
		\multicolumn{8}{c}{S3}                                                                                                          \\
		proposed       & 17.63(0.56)&1.19(1.10)&0.57(0.10)&15.04(1.37)&8.87(4.35)&0.77(0.13)&0.70(0.21)
		\\
		glinternet      & 13.78(2.12)   & 3.58(2.82)    & 1.35(0.15)      & 7.56(2.29)    & 2.26(1.79)    & 1.37(0.10)      & 1.99(0.54)    \\
		Lasso         & 6.74(2.14)    & 0.04(0.20)    & 1.73(0.13)      & 7.18(2.11)    & 7.18(4.05)    & 1.37(0.08)      & 2.47(0.72)    \\
		iFORM           & 16.04(2.45)   & 38.16(4.80)   & 1.44(0.40)      & 10.84(4.33)   & 31.98(3.71)   & 1.45(0.36)      & 2.35(1.16)    \\
		HierNet         & 10.36(2.05)   & 0.70(0.86)    & 1.48(0.12)      & 5.72(1.87)    & 5.68(3.68)    & 1.42(0.08)      & 2.21(0.66)    \\
		Grace              & 8.98(1.39)    & 0.20(0.40)    & 1.55(0.11)      & 10.08(1.41)   & 12.30(12.17)  & 1.28(0.08)      & 2.08(0.51)    \\
		GEL             & 17.88(0.63)   & 11.80(2.56)   & 0.77(0.15)      & 12.54(0.81)   & 134.22(25.84) & 2.13(0.30)      & 1.93(0.57)    \\ \midrule
		\multicolumn{8}{c}{S4}                                                                                                 \\
		proposed       & 17.94(0.24)&2.10(1.12)&0.37(0.07)&13.37(2.17)&15.10(3.98)&0.92(0.20)&0.77(0.20)
		\\
		glinternet      & 15.72(2.38)   & 4.06(3.13)    & 1.06(0.18)      & 7.48(3.21)    & 4.82(2.55)    & 1.44(0.09)      & 1.83(0.52)    \\
		Lasso         & 6.68(2.17)    & 0.00(0.00)    & 1.52(0.08)      & 3.62(2.28)    & 7.88(4.98)    & 1.54(0.07)      & 2.64(0.84)    \\
		iFORM           & 12.32(3.15)   & 43.94(5.57)   & 2.23(0.58)      & 3.80(4.55)    & 38.28(3.77)   & 2.17(0.41)      & 4.63(2.16)    \\
		HierNet         & 13.30(2.87)   & 1.10(1.45)    & 1.22(0.18)      & 6.50(2.88)    & 9.78(5.33)    & 1.62(0.13)      & 1.92(0.55)    \\
		Grace              & 9.50(2.46)    & 0.20(0.64)    & 1.45(0.13)      & 4.82(2.53)    & 11.42(8.43)   & 1.59(0.07)      & 2.35(0.73)    \\
		GEL             & 17.22(1.89)   & 8.98(4.75)    & 0.81(0.19)      & 11.78(2.39)   & 98.98(59.10)  & 1.90(0.51)      & 1.85(0.66)    \\ \bottomrule
	\end{tabular}
\end{table}

\begin{table}\label{t:two}
	\renewcommand\tabcolsep{1.4pt}
	\caption{Simulation results under the scenarios with $\rho=0.4, K=100$, and $r=1/\sqrt{12}$. In each cell, mean (SD) based on 100 replicates.}			
	\begin{tabular}{@{}lccccccc@{}}
		\toprule
		\textbf{Approach} & \textbf{M:TP} & \textbf{M:FP} & \textbf{M:RSSE} & \textbf{I:TP} & \textbf{I:FP} & \textbf{I:RSSE} & \textbf{PMSE} \\ \midrule
		\multicolumn{8}{c}{S1}                                                                                  \\
		proposed  & 16.92(0.85)&0.68(0.82)&0.42(0.08)&15.70(1.30)&3.44(2.17)&0.40(0.11)&0.60(0.15)
		\\
		glinternet & 15.16(2.68) & 3.46(2.88)  & 0.74(0.09) & 11.52(2.64) & 2.46(2.11)    & 0.78(0.07) & 1.14(0.33) \\
		Lasso    & 5.84(1.91)  & 0.00(0.00)  & 1.02(0.06) & 9.14(2.30)  & 11.12(5.36)   & 0.84(0.07) & 1.46(0.41) \\
		iFORM      & 11.42(2.12) & 44.50(3.45) & 1.59(0.19) & 5.40(2.17)  & 34.80(3.36)   & 1.38(0.12) & 2.57(0.79) \\
		HierNet    & 10.08(2.56) & 0.30(0.46)  & 0.92(0.08) & 5.96(2.13)  & 5.08(3.70)    & 0.93(0.06) & 1.48(0.35) \\
		Grace         & 7.06(2.05)  & 0.26(0.66)  & 1.01(0.08) & 10.10(1.96) & 5.92(5.60)    & 0.82(0.06) & 1.49(0.44) \\
		GEL        & 17.32(1.42) & 8.22(5.16)  & 0.69(0.12) & 11.88(1.75) & 92.68(60.45)  & 1.56(0.59) & 1.45(0.60) \\ \midrule
		\multicolumn{8}{c}{S2}                                                                                  \\
		proposed  & 15.92(1.54)&0.82(0.98)&0.46(0.09)&15.10(1.62)&3.40(2.36)&0.46(0.12)&0.65(0.19)
		\\
		glinternet & 14.40(3.28) & 3.66(2.85)  & 0.74(0.12) & 10.88(3.56) & 2.68(2.14)    & 0.81(0.10) & 1.16(0.36) \\
		Lasso    & 5.08(1.87)  & 0.12(0.48)  & 1.00(0.06) & 5.58(2.65)  & 40.72(95.41)  & 0.98(0.07) & 1.68(0.40) \\
		iFORM      & 11.02(2.30) & 44.94(4.67) & 1.62(0.23) & 4.74(2.14)  & 35.94(4.01)   & 1.43(0.13) & 2.52(0.84) \\
		HierNet    & 8.82(2.81)  & 0.40(0.76)  & 0.92(0.07) & 5.92(2.27)  & 4.72(3.30)    & 0.96(0.07) & 1.43(0.39) \\
		Grace         & 4.06(1.04)  & 0.14(0.40)  & 1.34(0.03) & 3.78(1.36)  & 5.48(5.59)    & 1.00(0.03) & 2.30(0.61) \\
		GEL        & 17.32(1.80) & 10.50(3.90) & 0.74(0.13) & 12.50(1.20) & 120.58(45.24) & 1.82(0.45) & 1.67(0.52) \\ \midrule
		\multicolumn{8}{c}{S3}                                                                                  \\
		proposed  & 16.17(1.44)&0.97(1.27)&0.58(0.08)&12.10(1.97)&4.83(3.70)&0.66(0.09)&0.69(0.18)
		\\
		glinternet & 10.64(2.46) & 1.88(1.78)  & 1.00(0.09) & 4.86(2.29)  & 0.88(1.33)    & 0.94(0.05) & 1.30(0.29) \\
		Lasso    & 5.32(1.56)  & 0.00(0.00)  & 1.21(0.10) & 5.34(1.76)  & 5.86(3.84)    & 0.93(0.04) & 1.51(0.40) \\
		iFORM      & 11.54(2.32) & 45.16(3.79) & 1.56(0.18) & 3.92(2.25)  & 36.58(3.41)   & 1.39(0.10) & 2.50(0.76) \\
		HierNet    & 8.38(2.02)  & 0.44(0.73)  & 1.06(0.08) & 3.98(2.05)  & 4.22(3.60)    & 0.96(0.05) & 1.38(0.36) \\
		Grace         & 7.22(1.56)  & 0.16(0.37)  & 1.13(0.09) & 8.02(1.32)  & 8.56(9.30)    & 0.87(0.04) & 1.35(0.35) \\
		GEL        & 17.62(1.58) & 11.40(2.67) & 0.73(0.12) & 11.76(1.49) & 131.62(31.37) & 1.91(0.30) & 1.68(0.47) \\ \midrule
		\multicolumn{8}{c}{S4}                                                                                  \\
		proposed  & 16.71(1.18)&1.06(1.06)&0.46(0.09)&9.15(2.44)&8.52(3.40)&0.83(0.11)&0.77(0.21)
		\\
		glinternet & 12.90(2.87) & 1.98(2.26)  & 0.82(0.09) & 3.68(2.11)  & 2.84(2.18)    & 0.98(0.04) & 1.19(0.20) \\
		Lasso    & 5.16(1.48)  & 0.00(0.00)  & 1.02(0.05) & 2.38(1.70)  & 5.44(3.72)    & 1.01(0.03) & 1.47(0.33) \\
		iFORM      & 10.12(1.89) & 46.70(3.32) & 1.88(0.19) & 1.22(1.45)  & 39.96(3.36)   & 1.65(0.12) & 3.11(0.82) \\
		HierNet    & 9.10(2.60)  & 0.36(0.53)  & 0.93(0.09) & 2.78(1.82)  & 5.42(3.94)    & 1.08(0.08) & 1.31(0.36) \\
		Grace         & 6.94(1.91)  & 0.12(0.48)  & 1.00(0.06) & 2.86(1.93)  & 7.78(7.08)    & 1.03(0.04) & 1.49(0.35) \\
		GEL        & 16.56(2.51) & 7.50(5.41)  & 0.71(0.13) & 10.76(2.50) & 83.72(65.78)  & 1.51(0.61) & 1.39(0.49) \\ \bottomrule
	\end{tabular}
\end{table}

It is observed that across the whole spectrum of simulation, the proposed approach has superior or similar performance compared to the alternatives with respect to both selection and prediction accuracy. It is able to identify the majority of true positives, while having much fewer false positives than most alternatives. For instance, under the scenario with setting S4 in Table 1, the proposed approach has (M:TP, M:FP, I:TP, I:FP) = (17.94, 2.10, 13.37, 15.10), compared to (15.72, 4.06, 7.48, 4.82) for glinternet, (6.68, 0.00, 3.62, 7.88) for Lasso, (12.32, 43.94, 3.80, 38.28) for iFORM, (13.30, 1.10, 6.50, 9.78) for HierNet, (9.50, 0.20, 4.82, 11.42) for Grace, and (17.22, 8.98, 11.78, 98.98) for GEL. Under the scenarios in Table 2 with lower signal level ($r=1/\sqrt{12}$), the advantages of the proposed approach become more prominent, especially under setting S4, where the important interactions have main effects with weaker signals. Specifically, the proposed approach has (M:TP, M:FP, I:TP, I:FP) = (16.71,	1.06, 9.15, 8.52), compared to (12.90,	1.98, 3.68,	2.84) for glinternet, (5.16, 0.00, 2.38,5.44) for Lasso, (10.12, 46.70, 1.22, 39.96) for iFORM, (9.10, 0.36, 2.78, 5.42) for HierNet, (6.94, 0.12, 2.86, 7.78) for Grace, and (16.56, 7.50, 10.76, 83.72) for GEL. The proposed approach also performs well in terms of estimation. For example, under setting S1 in Table 1, the proposed approach has (M:RSSE, I:RSSE)=(0.35, 0.45), compared to (0.96, 1.02) for glinternet, (1.48, 1.19) for Lasso, (1.35, 1.32) for iFORM, (1.23, 1.32) for HierNet, (1.46, 1.20) for Grace, and (0.76, 1.93) for GEL. In addition, higher prediction accuracy of the proposed approach is observed. For example, under  setting S2 in Table 2, the PMSEs are 0.65 (proposed), 1.16 (glinternet), 1.68 (Lasso), 2.52 (iFORM), 1.43 (HierNet), 2.30 (Grace), and 1.67 (GEL), respectively. Furthermore, we note that the proposed approach identifies all important networks correctly with N:FP=0 under all scenarios. In contrast, GEL cannot effectively select  important networks (especially under setting S4 with N:TP=2.47 and N:FP=0.06) and often misidentifies networks (details omitted). The glinternet approach generally has the second best performance, and under some scenarios with higher within network correlations ($\rho=0.6$) and simpler signal patterns (S1 and S2), it behaves competitively  in main-effect identification. However, the proposed approach can keep its superiority in interaction identification, estimation, and prediction. With a larger network size $(p_k=20)$, the proposed approach is again observed to perform favourably.

To mimic the scenarios under which a genetic factor (i.e. a gene) is involved in multiple networks, we conduct additional simulations with  $K=100$. Specifically, among the 1,000 genetic factors, there are 100 each of which  is involved in 2 to 6 networks. Summary results are provided in Tables S7-S10 (Section S2 of the Supplementary material). Similar conclusions can be drawn that the proposed approach has advantages over the alternatives.

\section{Data Analysis}

We analyze The Cancer Genome Atlas (TCGA) data on cutaneous melanoma (SKCM) and lung adenocarcinoma (LUAD) to identify important interactions (main effects, and networks) associated with phenotypes/outcomes. As one of the largest cancer genetics program, TCGA contains unique and valuable information. In this study, we consider mRNA gene expression measurements which are downloaded from the TCGA Provisional using the R package \textit{cgdsr}. Networks are constructed using the information from KEGG, which is a popular choice in recent network analysis studies \citep{Zhou2013Bayesian,Intro2013joint,Intro2019Liu}. Specifically, we follow \cite{Intro2019Liu} and obtain the network structures from KEGG pathway database using the R package \textit{KEGGgraph}, where each pathway is presented as a network with nodes being molecules (protein, compound, etc) and edges representing relation types between the nodes, e.g. activation or phosphorylation \citep{zhang2009kegggraph}. Here, we set ${A}^{(\boldsymbol{1})}_{k}(j,l)={A}^{(\boldsymbol{1})}_{k}(l,j)=1$ if the $j$th and $l$th genes are connected in the pathway and 0 otherwise.

\subsection{Cutaneous melanoma (SKCM) data}

The response of interest is the (log-transformed) Breslow's thickness, which is a measure of melanoma growth and has been widely used in the assessment of  melanoma. Data are available on 361 subjects and 19,904 gene expression measurements. Although the proposed approach is potentially applicable to a large number of genes, with the consideration that the number of cancer-related genes is not large, as well as to improve stability, a simple marginal screening is conducted. Specifically, the top 2,000 genes with the smallest p-values computed from a marginal linear model are selected. Among them, we construct 173 networks with the sizes from 1 to 50, containing 1,505 genes in total and 578 distinct genes after removing duplicates.

\begin{figure}
	\centering
	\includegraphics[width=4.5in]{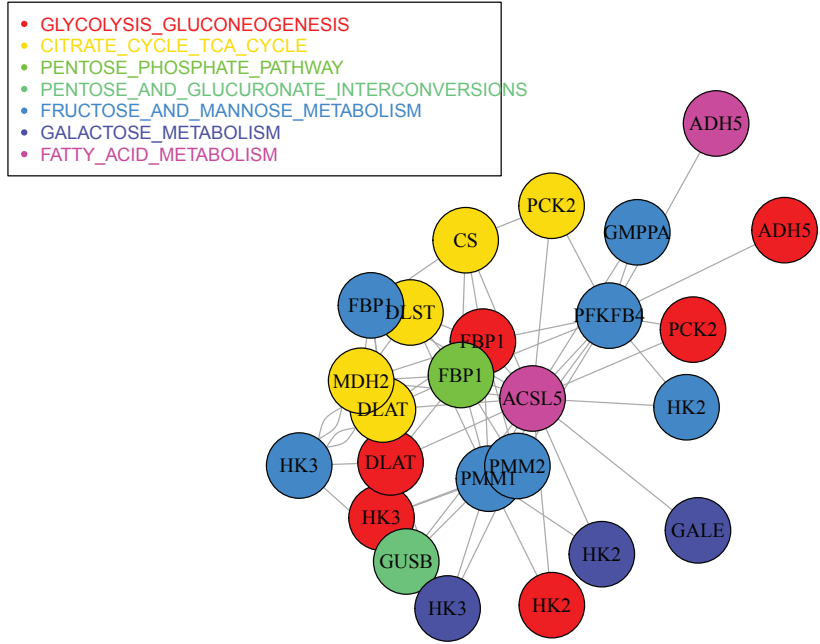}
	\caption{Analysis of the TCGA SKCM data using the proposed approach: identified main genetic
		effects, interactions, and networks. Different colors represent different networks, and two genes are connected if the corresponding interaction is also selected. We use different colors to represent genes that appear in different networks.}
	\label{fig:skcm}
\end{figure}

16 distinct main effects and 34 distinct interactions are identified by the proposed approach (25 main effects and 66 interactions before removing duplicates). The identified genes, their interactions, as well as networks are showed in Figure \ref{fig:skcm}, where two genes are connected if the corresponding interaction is also selected. The detailed estimation results are provided in Table S11 in the Supplementary material. Literature search suggests that the identified genes may be of great significance. For example, it has been found that high expression of gene PMM2 is associated with poor prognosis in melanoma. Gene FBP1, which is involved in three identified networks, has been shown to be significantly down-regulated in human melanoma cells. The expression of gene PCK2 has been found to be down-regulated in melanoma regenerative cells and closely related to the survival of tumor patients. Published studies have reported that gene PFKFB4, a known regulator of glycolysis, displays an unconventional role in melanoma cell migration and has increased expression levels in several human tumors including cutaneous melanoma. The simultaneous inactivation of genes HK1 and HK2 has been demonstrated to be sufficient to decrease the proliferation and viability of melanoma. In addition, gene PMM1 has been identified in published studies to be regulated in human melanoma and melanoma-associated pathways. We refer to Section S2 of the Supplementary material for the relevant references.

Furthermore, the proposed approach identifies seven networks, all of which are metabolics related and have important biological implications. For example, in a recent study, Citrate cycle (TCA cycle) has been suggested to be significantly down-regulated, while Galactose metabolism is up-regulated in tumor formation and progression. Other interesting networks have been linked to the development, progression, and outcome of melanoma. For instance, Fatty Acid metabolism has been shown to be essential for cancer cell proliferation. Glycolysis has been confirmed to play a significant role in developing metabolic symbiosis in metastatic melanoma progression. In addition, the Pentose phosphate has been found to be critical for cancer cell survival and ribonucleotide, as well as lipid biosynthesis.

Beyond the proposed approach, we also conduct analysis using the alternatives. The summary comparison results are presented in Table 3, where the numbers of main effects and interactions identified by different approaches, their overlaps, and RV coefficients are provided. Here, RV coefficient is a measure that describes the similarity of two matrices, and a larger value indicates a higher similarity. Different approaches are observed to identify quite different sets of main effects and interactions, and have moderate similarity as suggested by the RV coefficients. In particular, Lasso, Grace, and GEL, without accounting for the ``main effects-interactions'' hierarchy, identify a larger number of interactions than main effects. The other four approaches with the ``main effects-interactions'' hierarchy, including the proposed one, select a moderate number of main effects and interactions.

We further use a resampling approach to examine prediction performance and selection stability. The subjects are randomly partitioned into a training and a testing set. The mean PMSEs for the testing subjects over 100 resamplings are 0.57 (proposed), 0.60 (glinternet), 0.60 (Lasso), 1.16 (iFORM), 0.57 (HierNet), 0.64 (Grace), and 6.32 (GEL), suggesting the satisfactory prediction accuracy of the proposed approach. To evaluate selection stability, for each of the aforementioned important main effects and interactions, we compute its observed occurrence index (OOI), which is the selected frequency in 100 resamplings.
The proposed approach has the mean OOI value of 0.98, compared to 0.07 (glinternet), 0.28 (Lasso), 0.15 (iFORM), 0.62 (HierNet), 0.78 (Grace), and 0.50 (GEL). With the joint network selection and incorporated network structures, the proposed approach has significant improvement in selection stability. The prediction and stability analysis provides a certain degree of confidence to the proposed identification analysis.

\begin{table}\label{t:three}
	\renewcommand\tabcolsep{2pt}
	\caption{Data analysis: numbers of main effects and interactions (diagonal elements) identified by different approaches and their
		overlaps and RV coefficients (off-diagonal elements).}
	\begin{tabular}{cllllllll}
		\hline
		\multicolumn{1}{l}{\textbf{SKCM}}     &            &          &            &         &         &          &         &         \\
		\multicolumn{1}{l}{}                  &            & proposed & glinternet & Lasso   & iFORM   & HierNet  & Grace   & GEL     \\ \hline
		\multirow{7}{*}{\textbf{Main}}        & proposed   & 16       & 0(0.27)    & 0(0.03) & 0(0.42) & 0(0.40)  & 0(0.23) & 0(0.34) \\
		& glinternet &          & 15         & 1(0.22) & 5(0.58) & 10(0.62) & 0(0.11) & 0(0.41) \\
		& Lasso      &          &            & 1       & 1(0.12) & 0(0.03)  & 0(0.00) & 0(0.02) \\
		& iFORM      &          &            &         & 51      & 8(0.72)  & 0(0.19) & 1(0.67) \\
		& HierNet    &          &            &         &         & 51       & 1(0.33) & 1(0.59) \\
		& Grace      &          &            &         &         &          & 1       & 0(0.13) \\
		& GEL        &          &            &         &         &          &         & 28      \\ \hline
		\multirow{7}{*}{\textbf{Interaction}} & proposed   & 34       & 0(0.02)    & 0(0.00) & 0(0.08) & 0(0.00)  & 0(0.01) & 0(0.11) \\
		& glinternet &          & 9          & 6(0.59) & 1(0.08) & 1(0.40)  & 0(0.00) & 0(0.01) \\
		& Lasso      &          &            & 20      & 2(0.08) & 0(0.08)  & 0(0.01) & 0(0.02) \\
		& iFORM      &          &            &         & 44      & 0(0.16)  & 0(0.01) & 1(0.10) \\
		& HierNet    &          &            &         &         & 2        & 0(0.00) & 0(0.00) \\
		& Grace      &          &            &         &         &          & 16      & 1(0.03) \\
		& GEL        &          &            &         &         &          &         & 363     \\ \hline
		\multicolumn{1}{l}{\textbf{LUAD}}     &            &          &            &         &         &          &         &         \\
		\multicolumn{1}{l}{}                  &            & proposed & glinternet & Lasso   & iFORM   & HierNet  & Grace   & GEL     \\ \hline
		\multirow{7}{*}{\textbf{Main}}        & proposed   & 13       & 1(0.37)    & 0(0.24) & 0(0.41) & 1(0.45)  & 0(0.00) & 1(0.25) \\
		& glinternet &          & 16         & 1(0.49) & 8(0.71) & 13(0.78) & 0(0.00) & 0(0.33) \\
		& Lasso      &          &            & 4       & 2(0.57) & 1(0.54)  & 0(0.00) & 0(0.31) \\
		& iFORM      &          &            &         & 43      & 20(0.84) & 0(0.00) & 1(0.45) \\
		& HierNet    &          &            &         &         & 67       & 0(0.00) & 0(0.47) \\
		& Grace      &          &            &         &         &          & 0       & 0(0.00) \\
		& GEL        &          &            &         &         &          &         & 24      \\ \hline
		\multirow{7}{*}{\textbf{Interaction}} & proposed   & 38       & 0(0.07)    & 0(0.13) & 0(0.15) & 0(0.00)  & 0(0.02) & 0(0.12) \\
		& glinternet &          & 12         & 5(0.34) & 0(0.00) & 1(0.20)  & 0(0.00) & 0(0.03) \\
		& Lasso      &          &            & 253     & 4(0.12) & 3(0.22)  & 0(0.01) & 2(0.16) \\
		& iFORM      &          &            &         & 53      & 1(0.02)  & 0(0.00) & 0(0.06) \\
		& HierNet    &          &            &         &         & 13       & 0(0.00) & 0(0.02) \\
		& Grace      &          &            &         &         &          & 17      & 1(0.03) \\
		& GEL        &          &            &         &         &          &         & 269     \\ \hline
	\end{tabular}
\end{table}

\subsection{Lung adenocarcinoma (LUAD) data}

The response of interest is the reference value for the pre-bronchodilator forced expiratory volume in one second in percent (FEV1). It is a major indicator of pulmonary function impairment. Data are available on 232 subjects and 18,325 gene expression measurements. We conduct a similar prescreening, and 181 networks, containing 1,360 genes in total and 499 distinct genes are found.

With the proposed approach, 13 main effects and 38 interactions (13 main effects and 38 interactions before removing duplicates) are identified and presented in Figure \ref{fig:luad}. The detailed estimation results are provided in Table S12 in the Supplementary material. Independent evidences of their biological implications have been reported in the literature. For example, the ALDH2 locus has been associated with a higher risk of lung cancer among light smokers. Activated ACLY has been suggested as a negative prognostic factor in lung adenocarcinomas. Significantly higher ACSS2 expression has been observed in a substantial number of lung tumor samples. Published studies have demonstrated that late-stage LUAD patients have higher expression levels of HK2 and GBE1 than early-stage ones. Also, evidence suggests that the expression of PCK1 or PCK2 may be important for the growth of lung cancer due to high demand for anabolic metabolism and frequently insufficient supply. In addition, over-expression of PGAM1 has been observed in multiple human cancer types including lung cancer. References supporting the discussions
on biological functionalities are provided in Section S2 of the Supplementary material.

The proposed approach identifies two networks, which have been shown to have possible associations with lung cancer. For instance, studies have shown that as a reverse glycolysis pathway, gluconeogenesis can generate glucose from small carbohydrate precursors, which is crucial for the growth of tumor cells: the biosynthetic reaction in cancer cells is highly dependent on glycolysis intermediates. Recently, experiments on genetically engineered lung and pancreatic cancer tumors in mice have shown that the TCA cycle is highly affected by glucose metabolism, resulting in high intra-tumor and inter-tumor variability.
\begin{figure}
	\centering
	\includegraphics[width=4.5in]{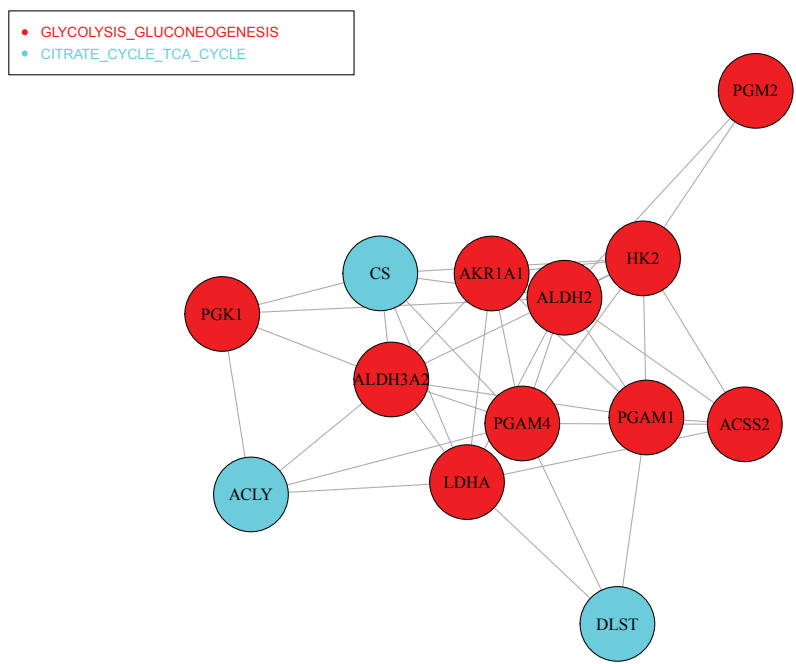}
	\caption{Analysis of the TCGA LUAD data using the proposed approach: identified main genetic
		effects, interactions, and networks. Different colors represent different networks, and two genes are connected if the corresponding interaction is also selected. We use different colors to represent genes that appear in different networks.}
	\label{fig:luad}
\end{figure}
We also conduct analysis using the alternatives and summarize the comparison results in Table 3. Similar to that for SKCM data, different approaches lead to identification results with low levels of overlapping. Prediction performance and selection stability are further examined based on 100 resamplings. The mean (PMSE, OOI) values are (0.05, 0.98) for the proposed approach, (0.05, 0.68) for glinternet, (0.06, 0.33) for Lasso, (0.22, 0.12) for iFORM, (0.04, 0.08) for HierNet, (0.04, 0.19) for Grace, and (0.73, 0.13) for GEL. The proposed approach again has competitive prediction accuracy and superior selection stability.

\section{Discussion}
\label{s:discuss}

In the study of complex diseases, gene-gene interaction analysis has attracted extended attention. Recently, biological networks have been accumulated, containing information on functionally related genetic groups and within-group structures. Thus, incorporating network information can potentially lead to a deeper biological understanding of phenotypes from a system perspective. In this study, we have developed the gene-gene interaction analysis, where the network information is incorporated. It advances from the existing interaction analyses by taking advantage of the assistance of network selection, where not only the ``main effects-interactions'' hierarchy but also the ``main effects/interations-networks'' hierarchy have been respected. In addition, motivated by the importance of link networks for interactions, the graph Laplacian Gaussian prior has been adopted to accommodate the underlying network structures of not only main effects but also interactions. As demonstrated in \cite{Intro2020laplace}, under certain regularization conditions, the graph Laplacian Gaussian prior has posterior consistency with a diverging number of nodes and edges in networks. The proposed approach may enjoy very broad applicability, where the networks can be potentially sparse or dense. The spike and slab priors for the regression coefficients and conjugate priors for the other parameters have been adopted, which offers the advantage of analytical simplification \citep{Narisetty2014Bayesian, gibbs}. Since results may be sensitive to the choice of hyperparameters $\zeta_j$'s, we have introduced a Beta prior on $\zeta_j$ to improve stability. The proposed approach can be formulated as a hybrid Bayesian model, with a solid statistical foundation and the potential to be effectively realized using the variational Bayesian expectation-maximization  algorithm. This is significantly advanced from the published Bayesian interaction analysis that usually adopts MCMC techniques, which have very low computational efficiency. Compared to some other variable selection techniques, such as penalization, Bayesian methods have demonstrated superiority in multiple aspects, such as providing readily available uncertainty estimates and a more informative approach to model selection \citep{vanerp2019shrinkage,gibbs}. Extensive simulation studies have been conducted, suggesting the practical superiority of the proposed approach in identification, estimation, and prediction. Two TCGA cancer studies have been used to illustrate application, leading to biologically sensible findings with satisfactory prediction accuracy and selection stability.

This study has focused on continuous response and assumed the Gaussian distribution. It can be of interest to extend the proposed approach to handle categorical and censored outcomes. For example, a data augmentation approach based on a probit model (categorical outcome) or an accelerated failure time model (censored outcome) can be potentially adopted \citep{Intro2011stingo}. However, our preliminary investigation suggests that this extension is expected to be nontrivial and warrants a separate work. The strong ``main effects-interactions'' hierarchy has been explored in this study, which is popular in recent interaction analysis \citep{Glin2015learning, hao2018model}. Modification can be potentially conducted in prior (\ref{equation:prior}) to respect the weak hierarchy. The duplication strategy has been adopted to accommodate overlappings in networks \citep{2009Group,Intro2013joint}, where we take the regression coefficients of the main effects involved in multiple networks as separate model parameters. As a result, if a main effect is selected, our model does not force all the networks this main effect is affiliated with to be selected. We acknowledge that when networks have a high overlapping level, the proposed analysis may not be stable. However, in practical data analysis with a moderate overlapping level, such as the SKCM data where some genes are involved in 30 to 50 networks, the proposed approach has been shown to be still validity. More prudent strategies are deferred to further investigation. In data analysis, we have utilized the KEGG information to construct networks. Other information sources, such as Gene Ontology terms and protein-protein interaction networks, can also be adopted.

\section*{Acknowledgements}

This work was supported by the National Institutes of Health [CA204120, CA121974, CA196530]; National Natural Science Foundation of China [12071273]; Bureau of Statistics of China [2018LD02]; ``Chenguang Program'' supported by Shanghai Education Development Foundation and Shanghai Municipal Education Commission [18CG42]; Program for Innovative Research Team of Shanghai University of Finance and Economics; Shanghai Pujiang Program [19PJ1403600]; and Fundamental Research
Funds for the Central Universities [2016110061, 2018110443, CXJJ-2019-413].

\bibliography{refajust}

\section*{SUPPLEMENTARY MATERIAL}

Supplementary material  may be found online in the Supplementary
material section at the end of the article.

\end{document}